
\input harvmac.tex
\input mssymb


\def\la{{\triangleright}}
\def\ra{{\triangleleft}}
\def\bicross{{\blacktriangleright\!\!\!\triangleleft}}
\def\cobicross{{\triangleright\!\!\!\blacktriangleleft}}

\def\id{{\bf 1}}
\def\Id{{1 \kern -.40em 1 }}
\def\half{{1 \over 2}}
\def\dotimes{{ \mathop{\otimes }^{\displaystyle .} }}
\def\vitimes{{ \mathop{,}^{\displaystyle \otimes} }}
\def\complex{{ {\rm C} \kern -.5em
   {\raise .12ex \hbox{\vrule height 1.2ex width 0.02em depth 0ex}}
   \kern 0.5em }}
\def\real{{ {\rm R} \kern -0.45em \vrule height 1.44ex width 0.02em depth 0ex
   \kern 0.45em }}

\def\Fqr{{ Fun(SO_q(N,\real)) }}
\def\FSO{{ Fun(SO(N-1,\real)) }}
\def\fSO{{ Fun(SO) }}

\def\Oqr{{ O_q^N(\real) }}
\def\FP{{ Fun({\cal P}_\gamma(N-1)) }}
\def\Ct{{ \widetilde{C} }}
\def\Ch{{ \widehat{C} }}
\def\R{{ \cal R }}
\def\order#1{{ O(R^{-#1}) }}
\def\podd{{ \mathop{+}^{\scriptscriptstyle odd} }}
\def\eps{{ \epsilon }}
\def\ot{{ \otimes }}
\def\oz{{ (0) }}
\def\ou{{ (1) }}
\def\od{{ (2) }}

\def\unpar{{ (1) }}
\def\deuxpar{{ (2) }}

\def\with{{ \qquad\hbox{with}\qquad }}
\def\and{{ \qquad\hbox{and}\qquad }}

\def\bcp{{bicrossproduct}}
\def\bu{{ \bar{u} }}
\def\bU{{ \bar{U} }}
\def\unbar{{ \bar{1} }}
\def\deuxbar{{ \bar{2} }}
\def\cP{{ \cal P}}
\def\cN{{ \cal N}}
\def\cPh{{ \hat{\cP} }}
\def\Mh{{ \hat{ M} }}
\def\Nh{{ \hat N }}

{\nopagenumbers
\rightline{MIT-CTP-2353}
\rightline{September 1994}
\rightline{hep-th/9409100}
\vskip 1in
\centerline{\titlefont The quantum Poincar\'e group from}
\smallskip
\centerline{\titlefont  quantum group contraction}
\bigskip
\centerline{Philippe Zaugg
\foot{Partially supported by the Swiss National Science Foundation
and by funds provided by the U.S. Department of Energy (D.O.E.) under
cooperative agreement \#DE-FC02-94ER40818.\hfill\break
} }
\bigskip
\centerline{\sl Center for Theoretical Physics}
\centerline{\sl Laboratory for Nuclear Science}
\centerline{\sl and Department of Physics}
\centerline{\sl Massachusetts Institute of Technology}
\centerline{\sl 77, Massachusetts Avenue}
\centerline{\sl Cambridge, MA 02139, USA}
\vskip 1in
\centerline{\bf Abstract}
\medskip
We propose a contraction of the de Sitter quantum group leading to the quantum
Poincar\'e group in any dimensions.
The method relies on the coaction of the de Sitter quantum group on a
non--commutative space, and the deformation parameter $q$ is sent to one.
The \bcp\ structure of the quantum Poincar\'e group is exhibited and shown to
be dual to the one of the $\kappa$--Poincar\'e Hopf algebra, at least in two
dimensions.
\vfill
\pageno=0
\eject}
\ftno=0

\newsec{Introduction}

In the realm of Hopf algebras, several propositions for a deformed enveloping
algebra of the $D$-dimensional Poincar\'e algebra $U_\kappa(\cP(D))$ have been
made in the recent past
\ref\italy{E.~Celeghini, R.~Giachetti, E.~Sorace and M.~Tarlini, J. Math. Phys.
{\bf 31} (1990) 2548; {\bf 32} (1991) 1155, 1159.}
\ref\lnr{J.~Lukierski, A.~Nowicki, H.~Ruegg and V.~Tolstoy, Phys. Lett. {\bf
B264} (1991) 331.\hfill\break
J.~Lukierski, A.~Nowicki, H.~Ruegg, Phys. Lett. {\bf B293} (1992)
344.}
\ref\lr{J.~Lukierski and H.~Ruegg, Phys. Lett. {\bf B329} (1994) 189.}
\ref\masla{P.~Maslanka, {\sl The $n$-dimensional $\kappa$--Poincar\'e algebra
and group}, Lodz University preprint.}.
The basic tool is a contraction of the simple de Sitter algebra $U_q(so(D+1))$,
in which the deformation parameter $q$ is sent to its classical value one.
One particular feature of these deformations is that they are minimal,
in the sense that the commutation relations are only slightly modified,
but not too minimal since they are not cocommutative any more.
Furthermore, these deformations are physically interesting since they involve
a dimensionful parameter $\kappa$ which sets a scale in the theory that could
in principle be determined by some measurement.

Mathematically, these Hopf algebras are deformations based on non semi--simple
Lie algebras. When the Lie algebra is simple, there is a natural dual Hopf
algebra, conventionnally known as the algebra of functions on the quantum
group, and the $R$--matrix provides the elegant link between these dual
structures
\ref\rtf{N.~Reshetikhin, L.~Takhtadzhyan and L.~Faddeev, Leningrad Math. J.,
Vol. {\bf 1} (1990), 193.}.
At the present time, there is no known $R$--matrix (except in dimension
three) for the deformed Poincar\'e algebra, therefore the investigation of the
dual along this line is impossible.
A potentially fruitful approach is to use the fact that the
$\kappa$--Poincar\'e algebra is an example of a \bcp\ of Hopf algebras, as was
recently shown in
\ref\majr{S.~Majid and H.~Ruegg, Phys. Lett. {\bf B334} (1994) 348.}.

Possible dual structures have been proposed by different authors, principally
obtained by the quantization of the Poisson bracket on the algebra of functions
on the classical group \lr\masla
\ref\zak{S.~Zakrzewski, J. Phys. A: Math. Gen. {\bf 27} (1994) 2075.}.

In this paper, we first extend a previous construction
\ref\joe{Ph.~Zaugg, {\sl The quantum 2D Poincar\'e group from quantum group
contraction}, MIT-CTP-2294 preprint, hepth-9404007.},
initially developed for
a two dimensional space--time, in which the quantum Poincar\'e group is
obtained by a contraction of the corresponding de Sitter quantum group (section
2, 3). The deformation parameter $q$ is sent to one as well, and similarly
a dimensionful parameter $\gamma$ enters the final algebra.
{}From a duality point of view, this is a natural starting point since the de
Sitter quantum group and the deformed enveloping de Sitter algebra are known to
be dual.

Next, by revealing the quantum Poincar\'e group bicrossproduct structure, we
provide a strong hint that this it is actually dual to the
$\kappa$--Poincar\'e algebra (sections 4, 5).
Finally in two dimensions, we are able to show that these bicrossproducts are
precisely dual to each other (section 6). This provides an alternative duality
proof to the one in
\ref\masb{P.~Maslanka, J. Math. Phys. {\bf 35} (1994) 1976.}.
Two appendices give some technical details used in the main text.

\newsec{The Hopf algebra $\Fqr$}

The complex orthogonal quantum group is defined in \rtf\
as the non--commutative algebra with unity and generators
$T=(t_{ij}), i,j=1,\dots,N$, subject to the relation $\R_t T_1 T_2 = T_2 T_1
\R_t$, where the $R$--matrix is
\eqn\Rt{
\eqalign{
 \R_t &= q \sum_{i \not= i'}^N e_{ii} \ot e_{ii} + \sum_{i,j;i \not= j,j'}^N
e_{ii} \ot e_{jj} + q^{-1} \sum_{i \not= i'}^N e_{i'i'} \ot e_{ii} \cr
 & + (q-q^{-1}) \sum_{i>j}^N
e_{ij} \ot e_{ji} - (q-q^{-1}) \sum_{i>j}^N q^{\rho_i -\rho_j} e_{ij} \ot
e_{i'j'} \podd e_{{N+1 \over 2},{N+1 \over 2}} \ot e_{{N+1 \over 2},
{N+1 \over 2}} .}
}
Here $\mathop{+}\limits^{\scriptscriptstyle odd}$
means that the term is present only for odd $N$. We use the
notation $i' = N+1-i$, the integer part $M=[{N-1 \over 2}]$ and the numbers
$\rho_i$, for $1 \leq i \leq M$,
$$
\rho_i = {N \over 2} - i, \qquad \rho_{i'} = -\rho_i, \qquad
\rho_{M+1} = 0 \quad\hbox{  (for odd $N$)}.
$$
The orthogonality conditions are
$$
T C T^T C^{-1} = C T^T C^{-1} T = \Id, \qquad \hbox{with} \qquad
C = \sum_{i=1}^N q^{\rho_i} e_{i'i}.
$$
The complete Hopf algebra structure is specified by the homomorphisms
\eqn\homos{
\Delta(T) = T \dotimes T, \qquad \eps(T) = \Id, \qquad
S(T) = C T^T C^{-1} .
}

The quantum $N$--dimensional complex space $O_q^N(\complex)$ is defined as the
non--commutative algebra with unity generated by the $N$ elements $x_i$ subject
to the relation
\eqn\xplane{
f(\hat{\R}_t) (x \ot x) = 0, \with
f(t) = {t^2 - (q+q^{1-N}) t + q^{2-N} \over q^{-1} + q^{1-N}} ,
}
and $\hat{\R}_t= P \R_t$ is the permuted $R$--matrix.
There is a coaction of the quantum group on the quantum space given by
\eqn\xcoaction{
\delta(x) = T \dotimes x,
}
which preserves the quadratic form $x^T C x$.

The quantum group real form we are considering here is specified by the
anti--involution
\eqn\realform{
T^* = D C^T T (C^{-1})^T D^{-1}
}
where $D={\rm diag}(\eps_1, \dots, \eps_N)$, with $\eps_i^2=1, \eps_{i'} =
\eps_i$ for $i=1,\dots,N$, and $\eps_i=1$ for $i=i'$. These $\eps$'s represent
in a way the signature of the quadratic form in the quantum space, and
characterize the real quantum algebra $Fun(SO_q(N,\real;\eps_i))$. Similarly
the quantum space is turned to a quantum real space $\Oqr$ with the help of
the anti--involution $x^* = D C^T x$.

For our geometric construction, it is more convenient to choose a real set
of generators for the quantum space, $z_i = M_{ij} x_j = z_i{}^*$,
with the matrix and its inverse
\eqn\matM{
\eqalign{
M &= {1 \over \sqrt{2}} \sum_{i=1}^N
(\alpha_i e_{ii} + \beta_i e_{i'i}) , \cr
M^{-1} &= {1 \over \sqrt{2}} \sum_{i=1}^N
(\gamma_i e_{ii} + \delta_i e_{i'i}) , \cr }
}
where
$$
\eqalign{
 (\alpha_1, \dots, \alpha_M) &= (1, \dots, 1) , \cr
 (\alpha_{M'}, \dots, \alpha_N) &= (-i\eps_{M} q^{\rho_M}, \dots,
-i\eps_{1} q^{\rho_1}) , \cr
 \beta_j = i \alpha_j, \qquad \gamma_j &= {1 \over \alpha_j}, \qquad \delta_j
= {1 \over \beta_{j'}} \qquad \hbox{for~} j\not= j' , \cr
 \alpha_{{N+1 \over 2}} =\beta_{{N+1 \over 2}} &=  \gamma_{{N+1 \over 2}} =
\delta_{{N+1 \over 2}} ={1 \over \sqrt{2} .}
}
$$
Accordingly, we take new real generators $V=(v_{ij})= M T M^{-1}$ for the
algebra $\Fqr$, which satisfy slightly different orthogonality conditions
\eqn\orthoz{
\eqalign{
V \Ch V^T &= \Ch, \qquad\quad \hbox{with} \qquad \Ch=M C M^T \cr
V^T \Ct V &= \Ct, \qquad\quad \hbox{with} \qquad \Ct=M^{-1T} C M^{-1} .}
}
The comultiplication and counit are similar to \homos, and the antipode is now
\eqn\realS{
S(V) = \Ch V^T \Ct .
}

In this real basis, the quantum space relations \xplane\ become
\eqn\qzplane{
\eqalign{
  z_i z_j - q z_j z_i - z_{i'} z_{j'} + q z_{j'} z_{i'} &= \hphantom{-}
i(z_{i'} z_j - q z_j z_{i'} + z_{i} z_{j'} - q z_{j'} z_{i}) \qquad
i<j,i<i',j<j' \cr
  z_{j'} z_{i'} - q z_{i'} z_{j'} - z_{j} z_{i} + q z_{i} z_{j} &=
-i ( z_{j'} z_{i} - q z_{i} z_{j'} + z_{j} z_{i'} -q z_{i'} z_{j}) \qquad
i<j,i>i',j<j' \cr
  z_{i'} z_{j} -q z_{j} z_{i'} + z_{i} z_{j'} -q z_{j'} z_{i} &=
-i( z_{i} z_{j} -q z_{j} z_{i}- z_{i'} z_{j'} +q z_{j'} z_{i'}) \qquad
i<j,i<i',j>j' \cr
  \eps_i[z_i , z_{i'}] = i {q^2 -1 \over q^2 +1} \sum_{k=i+1}^M
&\left({1+q^2 \over 2} \right)^{k-i} \eps_k (z_k^2+z_{k'}^2) \podd
i {q^2-1 \over q+1} \left( {1+q^2 \over 2} \right)^{M-i} z_{{N+1 \over 2}}^2 }
}
and the quadratic form is diagonal
\eqn\zmet{
z^T \Ct z = {1+q^{2-N} \over 1+q^2} \sum_{k=1}^M \left({1+q^2 \over 2}
\right)^{k} \eps_k (z_k^2+z_{k'}^2) \podd
{1+q^{2-N} \over 1+q} \left( {1+q^2 \over 2} \right)^{M} z_{{N+1 \over 2}}^2 .
}
In this equation, the meaning of $D={\rm diag}(\eps_1, \dots, \eps_N)$
as the signature of metric is clear, particularly in the limit $q \to 1$.

\newsec{Contraction}

We now apply the contraction procedure leading to the definition of the
quantum Poincar\'e group and the quantum space--time on which it coacts.
In the classical contraction scheme, the $N-1$ dimensional space--time is
identified with a neighbourhood of a particular point on the $N-1$ sphere (or
hyperbola if the signature is Minkowskian), in the limit of infinite radius.
Here we generalize this geometric point of view to non--commutative spaces.
The two dimensional situation was developed in details in \joe, both at the
classical and quantum level.

In the quantum space $\Oqr$, we consider a subspace of dimension $N-1$
characterized by the
condition $z^T \Ct z = const$ (this corresponds to
the de Sitter sphere in the classical Euclidean contraction).
This subspace is invariant under the quantum group coaction because the
quadratic form \zmet\ is invariant.
On this subspace, we select a particular point of coordinates
$(z_i)=(R,0,\dots,0)$ around which an expansion in $R$ is performed. In the
limit $R \to \infty$, this $N-1$ dimensional subspace will give rise to
the quantum space--time, and by a proper limit, the coaction \xcoaction\ will
induce a coaction of the quantum Poincar\'e group.

We consider elements of $\Oqr$ living on the subspace
\eqn\qsphere{
z^T \Ct z = \eps_1 {\cal R}^2 .
}
We absorb an irrelevant factor in $R^2 = 2{\cal R}^2/1+q^{2-N}$.
The factor $\epsilon_1$ is compulsory if we want to keep all coordinates real
when $R \to \infty$, as can be easily seen from \zmet\ (recall also that in
the contraction limit we choosed, $z_1 \to \infty$). The contraction amounts
to take simultaneously $R \to \infty$ and $q \to 1$ by letting
$q=\exp(\gamma/R)$, with $\gamma$ a finite constant.

In \qsphere\ we choose to expand $z_1$ as a series in $R$
\foot{Our convention for indices is that $i,j,k=1,\dots,N$, whereas
$a,b,c=2,\dots,N$.}
\eqn\zone{
z_1 = R \left( \id - {\eps_1 \over 2 R^2} \sum_{a=2}^N \eps_a z_a^2 +
\order{3} \right) .
}
Inserting this expansion in the relations \qzplane, the limit $R\to\infty$ is
well defined because all the divergent terms cancel, and we are left with the
unique constraint
\eqn\pplane{
[z_a, z_N] = -i \gamma z_a .
}
We therefore define the quantum space--time as the algebra generated by
the $z_a$ subject to the above constraint \pplane.

Next, we rewrite the generators of $\Fqr$ as an expansion in
the contraction parameter $R$
\eqn\vexp{
v_{ij} = \sum_{n=0}^\infty {v_{ij}^n \over R^n} ,
}
and from simple requirements we will collect enough informations on the
$v_{ij}^n$ to enable us to derive all the necessary relations characterizing
the algebra $\FP$.
First, we require that under the coaction $\delta$ of $\Fqr$, the elements
$z_{a}$ remain of order 1 in the limit $R \to \infty$. Since
$$
\delta(z_a) = v_{a1} \otimes z_1 + v_{ab} \otimes z_b ,
$$
and $z_1$ is of order $R$,
this is only possible if $v_{a1}^0 = 0$. Next we apply $\delta$ on both sides
of \zone\ to get
\eqn\deltaz{
{1 \over R}(v_{11} \otimes z_1 + v_{1a} \otimes z_a ) = \Id \otimes \id -
{\eps_1 \over 2 R^2} \sum_{a=2}^N \eps_a \delta(z_a^2) +
\order{3} ,
}
which implies that $v_{11}^0 = \Id$ and $v_{11}^1 \otimes \id + v_{1a}^0
\otimes z_a = 0$, since $\delta(z_a^2)$ are finite by construction.
As the elements $\id$ and $z_a$ are linearly independent, we also conclude
that $v_{1a}^0=0$ and $v_{11}^1=0$.

Collecting all this, we can take the $R \to \infty$ limit in $\delta(z) = V
\mathop{\otimes}\limits^{.} z$, and dividing $z_1$ by $R$, this yields
\eqn\Etrans{
\eqalign{
\delta(\id) &= \Id \ot \id ,\cr
\delta(z_a) &= v^1_{a1} \ot \id + v^0_{ab} \ot z_b .}
}
{}From this form of the coaction, we see that $v^1_{a1}$ play the role of
translations and $v^0_{ab}$ the role of Lorentz transformations.
It is then natural to take the elements $\Id, u_{ab} = v_{ab}^0$ and $u_a =
v_{a1}^1$ as the generators of $\FP$, the algebra of functions on the quantum
Poincar\'e group ${\cal P}_\gamma(N-1)$.

Now that we selected the generators of the algebra, we should determine the
constraints imposed on them by the previous quantum group structure.
First we apply the constraints that derive from the contraction of the
two orthogonality relations \orthoz.
At zeroth order in $1/R$ one gets respectively
\eqn\orthoa{
\eqalign{
v_{ab}^0 \eps_b v_{cb}^0 &= \eps_a \delta_{ac} , \cr
v_{ba}^0 \eps_b v_{bc}^0 &= \eps_a \delta_{ac} ,}
}
and at first order, the relations are
\eqn\orthob{
\eqalign{
  (v_{ji}^1 \eps_i v_{ki}^0 + v_{ji}^0 \eps_i v_{ki}^1 + \gamma v_{ji}^0
\eps_i \theta_i \rho_i v_{ki}^0 + i \gamma v_{ji}^0 \eps_i \rho_i v_{ki'}^0)
e_{jk} & = \gamma \eps_i \theta_i \rho_i e_{ii} + i \gamma \eps_i \rho_i
e_{ii'} \cr
  (v_{ij}^1 \eps_i v_{ik}^0 + v_{ij}^0 \eps_i v_{ik}^1 - \gamma v_{ij}^0
\eps_i \theta_i \rho_i v_{ik}^0 - i \gamma v_{ij}^0 \eps_i \rho_i v_{i'k}^0)
e_{jk} & = -\gamma \eps_i \theta_i \rho_i e_{ii} - i \gamma \eps_i \rho_i
e_{ii'} }
}
where $\theta_i = 1$ if $i \leq M$ and $\theta_i = -1$ if $i > M$.
These constraints will be useful when computing the antipode and
the commutation relations.

The next task is to determine the commutation relations among the generators
that derive from the contraction of the constraint $\R_v V_1 V_2 = V_2 V_1
\R_v$.  For that purpose, one need to expand that expression up to order
$R^{-2}$ (in order to include the relations of $v_{a1}^1$ with $v_{b1}^1$).
Higher order terms ($R^{-n}, n \geq 3$) will always contain elements $v_{ij}^n$
of that order which by definition do not belong to the quantum Poincar\'e
algebra, and thus do not yield new constraints on our set of generators.
Performing the expansion, we get for the first three terms
\eqna\rvvexp
$$
\eqalignno{
  [ V^\oz \vitimes V^\oz ] &= 0 , & \rvvexp a \cr
  [ V^\ou \vitimes V^\oz ] &+ [ V^\oz \vitimes V^\ou ] = [ \R^\ou, V^\oz
\otimes V^\oz ] , & \rvvexp b \cr
  [ V^\ou \vitimes V^\ou ] &= -[ V^\oz \vitimes V^\od ] - [ V^\od \vitimes
V^\oz ] - [ \R^\od, V^\oz \otimes V^\oz ] & \rvvexp c \cr
  &\hphantom{=~} - \R^\ou (V_1^\oz V_2^\ou + V_1^\ou V_2^\oz) +
(V_2^\oz V_1^\ou + V_2^\ou V_1^\oz) \R^\ou .}
$$
We used the shorthand notation $X = \sum_n X^{(n)} R^{-n}$ for all the
matrices, and the tensored commutator should be understood as $[V^{(n)}
\mathop{,}\limits^\ot V^{(m)} ]_{(ij,kl)} = [ v^n_{ik}, v^m_{jl} ]$.

Owing to the particular structure of $V^{(n)}$ obtained in \vexp--\Etrans,
equation {\rvvexp a} implies in components
\eqn\qpa{
[ u_{ab}, u_{cd} ] \equiv [ v^0_{ab}, v^0_{cd} ] = 0 .
}
{}From equation {\rvvexp b}, we extract the commutation relation between the
order zero and one generators of interest, $v^0_{cd}$ and $v^1_{a1}$, namely
we consider the component $(ac,1d)$ of that equation. The necessary elements
of the $R$--matrix are computed in Appendix A, and one gets the
commutation relations
\eqn\qpb{
[ u_a, u_{cd} ] \equiv [ v^1_{a1}, v^0_{cd} ] = i \gamma \left( (u_{Nd} -
\delta_{Nd}) \eps_1 \eps_a \delta_{ac} + (u_{cN} - \delta_{cN}) u_{ad}
\right) .
}

{}From equation {\rvvexp c}, we determine the commutation relation between the
order one generators, $v^1_{a1}$, considering the component $(ab,11)$. This
requires the knowledge of some particular matrix elements of $\R_v$ up to
order $R^{-2}$, which can be found in Appendix A. After some tedious but
straightforward algebra, the result is
\eqn\qpc{
[ u_a, u_b ] \equiv [ v^1_{a1}, v^1_{b1} ] = i \gamma ( \delta_{Na} u_b -
\delta_{Nb} u_a ) .
}
The other components of \rvvexp{}\ are not relevant since they involves
elements
which are not part of the quantum Poincar\'e algebra as defined after \Etrans.

The rest of the Hopf algebra structure is obtained by contracting the
comultiplication $\Delta(V) = V \mathop{\otimes}\limits^{.} V$, which yields:
\eqn\qpcom{
\Delta(u_{ab}) = u_{ac} \otimes u_{cb} ,
\qquad\qquad
\Delta(u_a) = u_a \otimes \id + u_{ab} \ot u_b ,
}
the counit $\eps(V) = \Id$:
\eqn\qpcounit{
\eps(u_{ab}) = \delta_{ab}
\qquad\qquad
\eps(u_a) = 0,
}
and the antipode \realS\
\eqn\qpS{
\eqalign{
  S(u_{ab}) &= \eps_a \eps_b  u_{ba} \qquad\qquad \hbox{no sum on $a,b$} \cr
  S(u_a) &= - \eps_a u_{ba} \eps_b u_b \qquad\; \hbox{no sum on $a$}. }
}

One readily check that the commutation relations \qpa--\qpc\ satisfy the
Jacobi identity and as they originate from a contraction of $\Fqr$, it is
natural to take them as the definition of the quantum Poincar\'e group $\FP$.
Furthermore this definition is consistent with previous ones \lr\masla\zak,
obtained from
quantization of a classical Poisson structure on the Poincar\'e group.

Looking closer at \orthoa\ and \qpa, one sees that $U=(u_{ab})$ actually
describes an ordinary orthogonal matrix (with commuting entries) which
preserves the metric $\eta_{ac} = \eps_1 \eps_a \delta_{ac}$, {\it i.e.}
\orthoa\ become $U^T \eta U = \eta$.
The introduction of the factor $\eps_1$ is suggested by \qsphere, \zone\ and
is natural when considering \qpb.
In particular, because of the constraint \realform\ imposing $\eps_1=\eps_N$,
in our construction the time has always a positive signature $\eta_{NN}=1$.
One should mention that for odd dimensional space--time, this also forces the
metric to have an odd/even numbers of plus/minus signs.

\newsec{The quantum Poincar\'e group as a \bcp\ of algebra}

It turns out that the Hopf algebra $\FP$ just constructed by contraction can
also be obtained as a \bcp\ of two Hopf algebras, whose general theory was
developed in
\ref\majid{S.~Majid, J. Algebra {\bf 130} (1990) 17.}.
Essentially, a \bcp\ is a way to build a non--commutative non--cocommutative
Hopf algebra from two Hopf algebras, using their respective (co)--actions on
one another, provided some conditions are satisfied. Appendix B summarizes this
construction.

In our case, the two algebras that form the \bcp\ are the algebra
of functions on the (classical) orthogonal group $Fun(SO(N-1,\real))$ and a
non--commutative deformation of the algebra of translation $T$. The algebra
$Fun(SO(N-1,\real))=A$  is generated as usual by the commuting elements $\bU =
(\bu_{ab})$
\foot{The indices $a,b,c,\dots$ takes their $N-1$ values from 2 to $N$,
in order to match with the notations in the previous sections.}
and has the Hopf algebra structure
\eqn\uso{
\eqalign{
\Delta (\bU ) &= \bU \dotimes \bU \cr
S(\bU) &= \eta^{-1} \bU^T \eta }
\qquad \qquad
\eqalign{
\eps( \bU ) &= \Id \cr
\bU^T \eta \bU &= \eta }
}
Recall that $\eta=\hbox{diag}(\eps_1 \eps_2, \dots, \eps_1
\eps_N)$ and represents the metric in $\real^{N-1}$.

The translation algebra $T = H$ is generated by the elements $\bu_{a}$ with
the following relations
\eqn\transl{
\eqalign{
[ \bu_a, \bu_b ] &= i \gamma (\delta_{aN} \bu_b - \delta_{bN} \bu_a) \cr
\Delta(\bu_a) &= \bu_a \ot \id + \id \ot \bu_a }
\qquad\qquad
\eqalign{
\eps(\bu_a) &= 0 \cr
S(\bu_a) &= - \bu_a }
}
$\fSO$ is a right $T$--module algebra with the structure map
$\alpha: \fSO \ot T\to \fSO$ given by
\eqn\stra{
\alpha(\bu_{ab} \ot \bu_c) \equiv  \bu_{ab} \ra \bu_c= i \gamma
\left( (\bu_{Nb} - \delta_{Nb})
\eta_{ac} + (\bu_{aN} - \delta_{aN}) \bu_{cb} \right) .
}
$T$ is a left $\fSO$--comodule coalgebra specified by the coaction
$\beta: T \to \fSO \ot T$
\eqn\strb{
\beta(\bu_a) = \bu_{ab} \ot \bu_b .
}
One can check that conditions (B.1) are satisfied by the structure
maps \stra, \strb,  therefore
$K=T \cobicross \fSO$ is a Hopf algebra. If in $K$ we denote the elements
$$
u_{ab} = \id \ot \bu_{ab},
\qquad\qquad
u_a = \bu_a \ot \id
$$
and we apply the definitions for the product of Appendix B, we
get the following relations in $K$
\eqn\relK{
\eqalign{
  [ u_{ab}, u_{cd} ] &= 0 \cr
  [ u_{a~}, u_{cd} ] &= i \gamma \left( (u_{Nd} - \delta_{Nd})
\eta_{ac} + (u_{cN} - \delta_{cN}) u_{ad} \right) \cr
  [ u_{a~}, u_{b~} ] &= i \gamma ( \delta_{Na}
u_b - \delta_{Nb} u_a ) }
}
and for the comultiplication, counit and antipode
\eqn\homoK{
\eqalign{
  \Delta(u_{ab}) &= u_{ac} \otimes u_{cb} \cr
  \eps(u_{ab}) &= \delta_{ab} \cr
  S(u_{ab}) &= \eta_{ac} u_{dc} \eta_{db} }
\qquad\qquad
\eqalign{
  \Delta(u_a) &= u_a \otimes \id + u_{ab} \ot u_b \cr
  \eps(u_a) &= 0 \cr
  S(u_a) &= - \eta_{ab} u_{cb} \eta_{cd} u_d }
}

This shows that the \bcp\ $K$ is in fact the quantum Poincar\'e group
built in the previous section, $\FP = T \cobicross \FSO$.

Ultimately we would like to show that this quantum Poincar\'e group is the Hopf
algebra dual to the $\kappa$--Poincar\'e algebra of \italy\lnr.
Since we established that the quantum Poincar\'e group is a left--right \bcp,
a first step in that direction is to verify that $\kappa$--Poincar\'e is a
right--left \bcp. This is done in the next section for the four dimensional
space--time.
To complete the duality proof, one should then prove that the action and
coaction of the quantum Poincar\'e group actually induce the coaction and
action of $\kappa$--Poincar\'e. This is technically difficult in general, and
for the time being we are able to perform it only for the two dimensional case.

\newsec{$\kappa$--Poincar\'e as a \bcp}

We explicitly construct the $\kappa$--Poincar\'e deformed algebra of
\italy\lnr\ as a left--right \bcp. A similar computation was proposed in \majr,
but for a right--left \bcp. The difference arises from our choice of the
opposite comultiplication for $\kappa$--Poincar\'e, a choice which is
arbitrary at the level of the Hopf algebra, but has the effect of permuting
left and right in the \bcp\ (co)--actions.
This time, the \bcp\ combines a deformation of the algebra of translations
with the enveloping algebra $U(so(3,1))$. The reduction to lower dimensional
space--time or to other metric signature is straightforward.

$T^*=B$ is a non--cocommutative deformation of the enveloping algebra of
translations with hermitian generators $\cP_\mu$ ($\mu,\nu=0,1,2,3$ and
$r,s,t=1,2,3$)
\eqn\translstar{
[ \cP_\mu , \cP_\nu ] =0, \qquad\qquad
\eqalign{
\Delta \cP_0 & = \cP_0 \ot \id + \id \ot \cP_0 , \cr
\Delta \cP_r & = \cP_r \ot e^{-\cP_0 /\kappa } + \id \ot \cP_r .}
}
$U(so(3,1))=G$ is simply the enveloping algebra of the Lorentz Lie
algebra, also with hermitian generators
\eqn\lolo{
[M_r,M_s] = i \eps_{rst} M_t, \qquad
[M_r,N_s] = i \eps_{rst} N_t, \qquad
[N_r,N_s] = -i \eps_{rst} M_t.
}
$T^*$ is turned into a left $U(so)$--module algebra by the action
\eqn\kPaction{
\eqalign{
  M_r \la \cP_0 & = 0, \qquad M_r \la \cP_s = i \eps_{rst} \cP_t,
\qquad N_r \la \cP_0 = i \cP_r, \cr
  N_r \la \cP_s & = i \delta_{rs} \left( {\kappa \over 2}(\id -
e^{-2 \cP_0 / \kappa}) + {1 \over 2 \kappa} \vec{\cP}^2 \right)
- {i \over \kappa} \cP_r \cP_s .}
}
$U(so)$ is a right $T^*$--comodule coalgebra with the coaction
\eqn\kPcoaction{
\eqalign{
  \delta(M_r) & = M_r \ot \id , \cr
  \delta(N_r) & = N_r \ot e^{-\cP_0/ \kappa} - {i \over \kappa} \eps_{rst}
M_s \ot \cP_t .}
}
These maps fulfill the conditions (B.2) and $L = T^* \bicross U(so(3,1))$
is a Hopf algebra. Putting
$$
\cPh_\mu = \cP_\mu \ot \id, \qquad \Mh_r = \id \ot M_r, \qquad
\Nh_r = \id \ot N_r
$$
one easily computes the following commutation relations in $L$:
\eqn\kPcr{
\eqalign{
  [ \cPh_\mu, & \cPh_\nu ] = 0, \qquad
[ \cPh_0, \Mh_r ] = 0, \qquad
[ \cPh_r, \Mh_s ] = i \eps_{rst} \cPh_t, \cr
 &  [ \Mh_r, \Mh_s ] = i \eps_{rst} \Mh_t, \qquad
[ \Nh_r, \Nh_s ] = -i \eps_{rst} \Mh_t, \cr
  [ \Nh_r, \cPh_0 ] =~ &  i \cPh_r, \qquad
[ \Nh_r, \cPh_s ] = i \delta_{rs} \left( {\kappa \over 2}(\id -
e^{-2 \cPh_0 / \kappa}) + {1 \over 2 \kappa} \vec{\cPh}^2 \right)
- {i \over \kappa} \cPh_r \cPh_s , }
}
and the comultiplications
\eqn\kPDelta{
\eqalign{
  \Delta(\cPh_0) & = \cPh_0 \ot \id + \id \ot \cPh_0, \qquad
\Delta(\cPh_r) = \cPh_r \ot e^{-\cPh_0 /\kappa} + \id \ot \cPh_r ,\cr
  \Delta(\Mh_r) & = \Mh_r \ot \id + \id \ot \Mh_r, \cr
  \Delta(\Nh_r) & = \Nh_r \ot e^{-\cPh_0 /\kappa} + \id \ot \Nh_r - {i \over
\kappa} \eps_{rst} \Mh_s \ot \cPh_t. }
}
The antipode follows also easily. As expected, the relations \kPcr\ and
\kPDelta\ are those of the $\kappa$--Poincar\'e Hopf algebra \italy\lnr, in a
somewhat different basis \majr.

\newsec{Duality in two dimensions}

We will prove the duality of the (co)--actions in the case of the two
dimensional Minkowski quantum Poincar\'e group. From \realform, the
restriction on the $\eps$'s imposes the signature $\eps_1 = \eps_3 = -1,
\eps_2=1$, therefore $\eta_{33}= 1 = -\eta_{22}$.
All the calculations below can be extended to the Euclidean situation without
difficulty.

In the Lorentz group, it is more convenient to describe a boost by the rapidity
parameter
\eqn\utheta{
\bU = (\bu_{ab}) = \left( \matrix{\cosh\theta & \sinh\theta \cr
\sinh\theta & \cosh\theta } \right) ,
}
and to take $\theta$ as the generator of $Fun(SO(1,1))$.

The algebra $T$ is generated by $\bu_2,\bu_3$ constrained by the commutator
$[\bu_3,\bu_2] =i \gamma \bu_2$ and a basis of it is given by the ordered
monomials $\bu_2^n \bu_3^m$ for non negative integers $n,m$.
{}From \stra\ and \utheta, the action of $T$ on $Fun(SO(1,1))$ is
$$
\eqalign{
  \theta \ra \bu_2 &= i \gamma( \id - \cosh\theta) ,\cr
  \theta \ra \bu_3 &= -i \gamma \sinh\theta ,}
$$
and the coaction \strb\ is
$$
\eqalign{
  \beta(\bu_2) &= \cosh\theta \ot \bu_2 + \sinh\theta \ot \bu_3 ,\cr
  \beta(\bu_3) &= \sinh\theta \ot \bu_2 + \cosh\theta \ot \bu_3 .}
$$
In $T^*$ we take the commuting generators $P_2,P_3$ with the pairing
$$
\eqalign{
\vev{\bu_2^n \bu_3^m,P_2} &= \delta_{n,1} \delta_{m,0} \cr
\vev{\bu_2^n \bu_3^m,P_3} &= \delta_{n,0} \delta_{m,1} }
\qquad\Rightarrow\qquad
\vev{\bu_2^n \bu_3^m,P_2^p P_3^q} = n!m!\delta_{n,p} \delta_{m,q} ,
$$
and from the $\bu_a$'s commutation relation we deduce their comultiplication
$$
\eqalign{
\Delta(P_2) &= P_2 \ot \id + e^{i\gamma P_3} \ot P_2 ,\cr
\Delta(P_3) &= P_3 \ot \id + \id \ot P_3 .}
$$

In $Fun(SO(1,1))^*=U(so(1,1))$ we single out the generator $N$
with the pairing
$$
\vev{\theta^n,N} = \delta_{n,1}.
$$
Since $Fun(SO(1,1))$ is a right $T$--module, $U(so(1,1))$ should be a right
$T^*$--comodule. The most general coaction is
$$
\delta(N) = c_{p,nm} N^p \ot P_2^n P_3^m.
$$
To compute the coefficients, we use the duality relation
$$
\vev{\theta^p \ra \bu_2^n \bu_3^m, N} =
\vev{\theta^p \ot \bu_2^n \bu_3^m, \delta(N)} = p! n! m! c_{p,nm} .
$$
{}From
$$
p! c_{p,1\;0} = \vev{\theta^p \ra \bu_2,N} = \vev{p\theta^{p-1} i \gamma
(1-\cosh\theta),N} = 0,
$$
we deduce that
\eqn\cpnzero{
p! n! c_{p,n\;0} = \vev{\theta^p \ra \bu_2^n,N} =
\vev{(\theta^p \ra \bu_2^{n-1}) \ot \bu_2 ,\delta(N)} =0 .
}
Similarly, one easily finds that $c_{p,0\;1} = -i\gamma\delta_{p,1}$,
which allows to establish the recurrence relation
$$
p! m! c_{p,0\;m} = \vev{(\theta^p \ra \bu_3^{m-1}) \ot \bu_3, \delta(N)}
= -i\gamma \vev{\theta^p \ra \bu_3^{m-1},N}
= -i \gamma p! (m-1)! c_{p,0\;m-1},
$$
solved by
\eqn\cpzerom{
c_{p,0\;m} = {(-i\gamma)^m \over m!} \delta_{p,1} .
}
The coefficients for strictly positive $n,m$ vanish since
$$
\vev{\theta^p \ra \bu_2^n \bu_3^m, N} =
\vev{(\theta^p \ra \bu_2^n) \ot \bu_3^m, \delta(N)}=
\vev{\theta^p \ra \bu_2^n, N} \vev{\bu_3^m, e^{-i\gamma P_3}} = 0 ,
$$
as a consequence of \cpnzero\ and \cpzerom. Therefore the coaction is
\eqn\deltaco{
\delta(N) = N \ot e^{-i \gamma P_3} .
}

As $T$ is a left $Fun(SO(1,1))$--comodule, $T^*$ should be a left
$U(so(1,1))$--module, and we have to compute
$$
N \la P_a = d_{a,nm} P_2^n P_3^m ,
$$
using the duality
$$
\vev{\bu_2^n \bu_3^m, N \la P_a} = \vev{\beta(\bu_2^n \bu_3^m),
N \ot P_a} = n! m! d_{a,nm} .
$$
For $U$ any element of $T^*$, we have from (B.1)
\eqn\betarec{
\beta(U \bu_a) = U^\unbar \ra \bu_a \ot U^\deuxbar + U^\unbar \bu_{ac} \ot
U^\deuxbar \bu_c .
}
Therefore, using the coaction \deltaco, we get
\eqn\dcoef{
\vev{\beta(U \bu_a), N \ot P_b} = -i\gamma \delta_{a,3} \vev{\beta(U),
N \ot P_b} + \vev{U^\unbar u_{ac},N} \vev{U^\deuxbar \ot u_c,\Delta(P_b)}.
}
For $b=3$, the second term always vanishes except when $U=\id$ and $a=2$ and
we find
$$
\vev{\beta(\bu_2^n \bu_3^m),N \ot P_3} = (-i\gamma)^m \delta_{n,1} =
n! m! d_{3,nm},
$$
which yields
\eqn\nptrois{
N \la P_3 = P_2 e^{-i\gamma P_3} .
}

When the index $b=2$, \dcoef\ reduces to
$$
\vev{\beta(U \bu_a), N \ot P_2} = \delta_{a,3} \left(-i\gamma \vev{\beta(U),
N \ot P_2} + \vev{\beta(U), 1 \ot e^{i \gamma P_3}} \right)
+ \delta_{a,2} \vev{\beta(U),N \ot e^{i \gamma P_3}} .
$$
Using the pairings
$$
\eqalign{
  \vev{\beta(\bu_2^n \bu_3^m), \id \ot e^{i \gamma P_3}} &= (i\gamma)^m
\delta_{n,0} \cr
  \vev{\beta(\bu_2^n \bu_3^m), N \ot e^{i \gamma P_3}} &= i\gamma
\delta_{n,1} \delta_{m,0} }
$$
we get, for $n>0$
$$
\vev{\beta(\bu_2^n \bu_3^m),N \ot P_2} = (-i \gamma)^m (i\gamma)
\delta_{n-1,1} = n! m! d_{2,nm} ,
$$
and for $n=0$
$$
\vev{\beta(\bu_3^m),N \ot P_2} = -i \gamma
\vev{\beta(\bu_3^{m-1}),N \ot P_2} + (i\gamma)^{m-1} .
$$
This last recurrence is solved by the coefficients
$$
d_{2,0\;2m} = 0, \qquad\qquad
d_{2,0\;2m+1} = {1 \over i\gamma} {(i \gamma)^{2m+1} \over (2m+1)!},
$$
and we finally get the action
\eqn\npdeux{
N \la P_2 = {1 \over i\gamma}\sinh(i\gamma P_3) + {i \gamma \over 2} P_2^2
e^{-i\gamma P_3} .
}

Before making contact with the previous section, we should be careful about
the hermitian properties of the generators $N,P_a$. Knowing that $\theta,
\bu_a$ are hermitian, these are established using the definition (see
\ref\tom{T.~Koornwinder, {\sl General compact quantum groups, a tutorial},
University of Amsterdam Math. preprint 94-06, hepth-9401114.}
for example)
$$
\vev{\bu_2^n \bu_3^m,P_a^*} = \overline{ \vev{S(\bu_2^n \bu_3^m)^*,P_a} },
\qquad
\vev{\theta^n,N^*} = \overline{ \vev{S(\theta^n)^*,N} },
$$
and we find that $N,P_a$ are actually anti--hermitian.

If we define the new hermitian
generators $\cN = i N, \cP_2 = i P_2 e^{-i\gamma P_3}, \cP_3=i P_3$,
\deltaco, \nptrois\ and \npdeux\ become
\eqn\bla{
\eqalign{
  \delta(\cN) &= \cN \ot e^{-\gamma \cP_3} ,\cr
  \cN \la \cP_3 &= i \cP_2 ,\cr
  \cN \la \cP_2 &= {i \over 2 \gamma}\left(\id - e^{-2\gamma \cP_3}\right)
-{i\gamma \over 2} \cP_2^2 ,}
}
which is clearly the reduction of the maps \kPaction\ and \kPcoaction\
to the two dimensional situation, with the substitution $\kappa = 1/\gamma$.
Therefore the two dimensional $\kappa$--Poincar\'e is dual to the quantum
Poincar\'e group.

\newsec{Conclusion}

There are good reasons to believe that the quantum Poincar\'e group is in fact
the dual to the $\kappa$--Poincar\'e Hopf algebra. The approach proposed here
is very reminiscent of the contraction used in deriving $\kappa$--Poincar\'e:
we start from a dual structure and the deformation parameter $q$ is treated in
the same way. Furthermore the \bcp\ formulation of these two Hopf algebras
appear to be dual to each other, as the two dimensional proof of section 6
shows it.

The advantage of using the \bcp\ structure of the quantum Poincar\'e group and
algebra is that they are split into their building blocks which are easier to
handle, being simpler mathematical structures. The algebra of functions on the
classical Lorentz group is dual to the envelopping algebra of the Lorentz Lie
algebra
\ref\hel{S.~Helgason, {\sl Differential geometry, Lie groups and symmetric
spaces}, Academic Press, London 1978.}
and obviously $T^*$ is dual to $T$. Therefore, as vector spaces,
$\kappa$--Poincar\'e and the quantum Poincar\'e group are dual.
Showing that the algebraic structures on these spaces are dual reduces to the
proof of the (co)--actions duality.

The difficulty in generalising  the result of section 6 to higher
dimensions lies mainly in the definition of dual basis.
The presentation of the {\bcp}s are simpler in the respective basis
\uso--\strb\ and \translstar--\kPcoaction, but these are very unconvenient
basis when dealing with the duality issue.

\appendix{A}{The $R$--matrix}

In order to derive the commutation relations \rvvexp{}\ of the quantum
Poincar\'e group by contraction, we need to expand the $R$--matrix \Rt\ up to
second order in $R$.
First, one has to express it in the $v_{ij}$ basis, using the matrix $M$ \matM
$$
\R_v = M \ot M \; \R_t \; M^{-1} \ot M^{-1} = \sum_{n=0}^\infty \R_v^{(n)}
R^{-n}.
$$
It is obvious that the zeroth order term $\R_v^\oz$
is the identity matrix, and this explains the simplicity
of the result \qpa.

The first order term $\R_v^\ou$ can be recast after some algebra in the
conventional form
\eqn\Run{
\R_v^\ou = \gamma \sum_{i=1}^M H_i \ot H_i + 2 \gamma \sum_{\alpha \in
\Delta_+} E_{-\alpha}  \ot E_{\alpha} .
}
In that equation, $H_i, E_{\pm \alpha}$ are the Cartan--Weyl generators of the
Lie algebra $so(N,\eps)$ in the defining representation. Given an orthonormal
basis $e_i, 1 \leq i \leq M$ of $\real^M$, the positive (long) roots
$\Delta_+$ are $e_i \pm e_j, 1 \leq i < j \leq M$, and when $N$ is odd, the
additional positive short roots are $e_i, 1 \leq i \leq M$. Putting
$$
N_{ij} =  -i \eps_i e_{ij} + i \eps_j e_{ji} ,
$$
these generators are
\eqn\generators{
\eqalign{
H_i &= \eps_i N_{ii'} \cr
  E_{e_i \pm e_j} &= \half ( N_{ij} + i N_{i'j} \pm i N_{ij'} \mp N_{i'j'} )
\qquad\quad
E_{e_i} = {1 \over \sqrt{2}} ( N_{i{N+1 \over 2}} + i N_{i'{N+1 \over 2}} )
\cr
  E_{-(e_i \pm e_j)} &= {\eps_i \eps_j \over 2} ( N_{ij} - i N_{i'j} \mp i
N_{ij'} \mp N_{i'j'} )
\qquad
E_{-e_i} = {\eps_i \over \sqrt{2}} ( N_{i{N+1 \over 2}} - i N_{i'{N+1 \over
2}} ) }
}

For the second commutator $[ v_{a1}^1 , v_{cd}^0 ]$ one first remarks that
due to the structure of the matrices $V^{(0,1)}$, the term
$[ V^\oz \mathop{,}\limits^\ot V^\ou ]_{(ac,1d)}$ vanishes, and
the right hand side is
$$
[ \R^\ou, V^\oz \otimes V^\oz ]_{(ac,1d)} = \R^\ou_{(ac,1b)} v_{bd}^0 -
v_{ab}^0 v_{ce}^0 \R^\ou_{(be,1d)} .
$$
{}From \Run\ and \generators\ one computes the relevant matrix element
$$
\R^\ou_{(ab,1c)} = i \gamma (  \delta_{Nb} \delta_{ac}
- \eps_1 \eps_a \delta_{Nc} \delta_{ab} ) ,
$$
and we get
$$
[ \R^\ou, V^\oz \otimes V^\oz ]_{(ac,1d)} = - i \gamma \left( (v^0_{Nd} -
\delta_{Nd}) \eps_1 \eps_a \delta_{ac} + (v^0_{cN} - \delta_{cN}) v^0_{ad}
\right) .
$$

For the last commutator $[ v_{a1}^1 , v_{b1}^1 ]$, the terms $[ V^\oz
\mathop{,}\limits^\ot V^\od ]_{(ab,11)}$ and $[ V^\od \mathop{,}\limits^\ot
V^\oz ]_{(ab,11)}$ vanish. For the rest, we need the linear and quadratic
terms in $\R_v$ and in particular
\eqn\appr{
\left( \R^\ou (V_1^\oz V_2^\ou + V_1^\ou V_2^\oz) +
 (V_2^\oz V_1^\ou + V_2^\ou V_1^\oz) \R^\ou \right)_{(ab,11)} .
}
Again, due to the specific structure of the $V^{(0,1)}$ matrices, only some
additional matrix elements enter the above equation and are found to be
$$
\eqalign{
  \R^\ou_{(ab,c1)} &= -i \gamma (  \delta_{Na} \delta_{bc}
- \eps_1 \eps_a \delta_{Nc} \delta_{ab} ) , \cr
  \R^\ou_{(ic,11)} &= -\gamma \eps_1 \eps_i \delta_{ic} = \R^\ou_{(ci,11)} .}
$$
When inserted in \appr, one gets
\eqn\termfour{
\eqalign{
  \left( \R^\ou (V_1^\oz V_2^\ou + V_1^\ou V_2^\oz) \right)_{(ab,11)} &=
i \gamma (\delta_{Nb} v^1_{a1} - \delta_{Na} v^1_{b1}) , \cr
  \left( V_2^\oz V_1^\ou + V_2^\ou V_1^\oz) \R^\ou \right)_{(ab,11)} &=
- \gamma \eps_1 \eps_c ( v^0_{bc} v^1_{ac}  + v^1_{bc} v^0_{ac} ) .}
}
The last contribution to {\rvvexp c} is
\eqn\termfive{
[ \R^\od, V^\oz \otimes V^\oz ]_{(ab,11)} = \R^\od_{ab,11} - v^0_{ac} v^0_{bd}
\R^\od_{cd,11} .
}
Expanding the $R$--matrix up to order two, one gets
$$
\R^\od_{ab,11} = \gamma^2 (-\eps_1 \eps_a \theta_a \rho_a \delta_{ab} -
i\eps_1 \eps_b \rho_b \delta_{ab'} + 2 \eps_1 \eps_a \rho_1 \delta_{ab} ) .
$$
Therefore \termfive\ becomes (sum on $c$ only)
$$
\eqalign{
  [ \R^\od, V^\oz \otimes V^\oz ]_{(ab,11)} &= \gamma^2 \eps_1 ( v^0_{bc}
\eps_c \theta_c \rho_c v^0_{ac} + i v^0_{bc} \eps_c \rho_c v^0_{ac'} - \eps_a
\theta_a \rho_a \delta_{ab} - i \eps_b \rho_b \delta_{ab'} ) \cr
  &= -\gamma \eps_1 \eps_c ( v^0_{bc} v^1_{ac}  + v^1_{bc} v^0_{ac} ) . }
$$
In the last step, we used the first orthogonality relation \orthob\ (putting
$j=b,k=a$) in order to simplify the expression, and one sees that it cancels
against the second contribution in \termfour, leaving the result \qpc.

\appendix{B}{The \bcp}
In this appendix, we recall the \bcp\ construction of Majid, setting up the
notations that are used in the main text. Some detailed proofs can be found
in \majid.

Let $H,A$ be two Hopf algebras, where $A$ is a right $H$--module algebra with
the structure map $\alpha: A \ot H \to A$,
$$
\alpha(a \ot h) = a \ra h \qquad\qquad h \in H, a \in A,
$$
and $H$ is a left $A$--comodule coalgebra with structure map
$\beta: H \to A \ot H$,
$$
\beta(h)=  h^\unbar \otimes h^\deuxbar
\qquad\qquad h,h^\deuxbar \in H, h^\unbar \in A.
$$
On the smash product--coproduct $K = H \# A$
(which is isomorphic to $H \ot A$ as a
vector space) one can put both a structure of algebra with the
multiplication rule
$$
(h \ot a) \cdot (g \ot b) = h g_\unpar \ot (a \ra g_\deuxpar) b ,
$$
and a structure of coalgebra with the comultiplication
$$
\Delta(h \ot a) = h_\unpar \ot h_\deuxpar{}^\unbar a_\unpar \ot
h_\deuxpar{}^\deuxbar \ot a_\deuxpar .
$$
The comultiplication is denoted by $\Delta(h)= h_\unpar \ot h_\deuxpar$.
$K$ is a bialgebra if and only if
\eqn\appa{
\eqalign{
  \eps(a \ra h) = \eps(a) \eps(h) &\and
\beta(1) = 1 \ot 1 \cr
  \Delta(a \ra h) & = (a_\unpar \ra h_\unpar) h_\deuxpar{}^\unbar \ot
a_\deuxpar \ra h_\deuxpar{}^\deuxbar \cr
  \beta(hg) & = (h^\unbar \ra g_\unpar) g_\deuxpar{}^\unbar \ot
h^\deuxbar g_\deuxpar{}^\deuxbar \cr
  h_\unpar{}^\unbar (a \ra h_\deuxpar ) \ot h_\unpar{}^\deuxbar & =
(a \ra h_\unpar) h_\deuxpar{}^\unbar \ot h_\deuxpar{}^\deuxbar }
}
These conditions arise from the compatibility of the multiplication and the
comultiplication in $K$.
Then $K$ is even a Hopf algebra with the antipode
$$
S(h \ot a) = (1 \ot S(h^\unbar a)) \cdot (S(h^\deuxbar) \ot 1) .
$$
$K$ is called a right--left \bcp\ and is denoted by $H \cobicross A$.

This structure has a nice dual counterpart, where left and right get
exchanged. Let this time
$B$ be a left $G$--module algebra with
the structure map $\gamma: G \ot B \to B$,
$$
\gamma( g \ot b ) = g \la b \qquad\qquad g \in G, b \in B,
$$
and $G$ be a right $B$--comodule coalgebra with structure map
$\delta: G \to G \ot B$,
$$
\delta(g)=  g^\unbar \otimes g^\deuxbar \qquad\qquad g,g^\unbar \in G,
g^\deuxbar \in B.
$$
On the smash product--coproduct $L = B \# G$ the multiplication rule is
$$
(a \ot h) \cdot (b \ot g) = a (h_\unpar \la b) \ot h_\deuxpar g ,
$$
and the comultiplication
$$
\Delta( b \ot g) = b_\unpar \ot g_\unpar{}^\unbar \ot b_\deuxpar
g_\unpar{}^\deuxbar \ot g_\deuxpar .
$$
$L$ is a bialgebra iff
\eqn\appb{
\eqalign{
  \eps(g \la b) = \eps(g) \eps(b) &\and
\delta(1) = 1 \ot 1 \cr
  \Delta( g \la b) & =  g_\unpar{}^\unbar \la b_\unpar \ot g_\unpar{}^\deuxbar
(g_\deuxpar \la b_\deuxpar) \cr
  \delta (gh) & =  g_\unpar{}^\unbar h^\unbar \ot g_\unpar{}^\deuxbar
(g_\deuxpar \la h^\deuxbar) \cr
  g_\deuxpar{}^\unbar \ot (g_\unpar \la b) g_\deuxpar{}^\deuxbar &=
g_\unpar{}^\unbar \ot g_\unpar{}^\deuxbar (g_\deuxpar \la b) }
}
Then $L$ is a Hopf algebra with antipode
$$
S(b \ot g) = (1 \ot S(g^\unbar)) \cdot ( S(b g^\deuxbar) \ot 1 ) ,
$$
and is called a left--right \bcp\ denoted by $B \bicross G$.

\listrefs

\bye